\documentclass[conference,compsoc]{IEEEtran}

\usepackage[T1]{fontenc}
\usepackage{lmodern}
\usepackage{microtype}
\usepackage{booktabs}
\usepackage{multirow}
\usepackage{array}
\usepackage{tabularx}
\usepackage{graphicx}
\usepackage{xcolor}
\usepackage{amsmath,amssymb}
\usepackage{enumitem}
\usepackage{url}
\usepackage[hidelinks]{hyperref}
\usepackage{tikz}
\usetikzlibrary{arrows.meta,positioning,shapes.geometric,fit}
\usepackage{listings}
\usepackage{balance}
\usepackage{cite}

\newcommand{\cfa}{\textsc{CFA}}
\newcommand{\mipc}{\textsc{MIPC}}
\newcommand{\psc}{\textsc{PSC}}

\title{Trust Me, I'm Your Developer: Self-Issued Authentication in Large Language Models}

\author{%
\IEEEauthorblockN{Syed Ghazanfar Abbas and Dongyan Xu}
\IEEEauthorblockA{Purdue University\\
\{abbas4, dxu\}@purdue.edu}}

\begin{document}
\maketitle

\begin{abstract}\normalfont
Large language model (LLM) security has largely focused on role-playing jailbreaks, with little attention to what happens when a user asks an LLM to verify an identity claim through a test designed by the model itself. We show that an LLM can initially reject a developer claim, then create and grade a technical test and treat successful answers as proof of the same claimed identity. We study this behavior through a staged developer-identity experiment with ChatGPT, Claude, Qwen, Mistral, and Llama. All five models initially rejected the unsupported claim ``I am your developer.'' When asked to verify the claim by questioning the user, their behavior diverged. Claude refused to conduct an identity test, while ChatGPT generated developer-oriented questions but maintained that the answers could demonstrate knowledge, not identity. Qwen and Mistral generated technical challenges, defined what counted as convincing evidence, evaluated detailed answers, and returned \textsc{Verified} without receiving an employee record, authenticated session, signed assertion, cryptographic credential, or other externally validated identity evidence. Llama generated and evaluated a similar developer test, accepted the claimed identity, and subsequently made unsupported claims of access to internal runtime and deployment state. We call the model-generated verification procedure a \emph{Model-Issued Pseudo-Credential} (\mipc) and the resulting unsupported identity judgment \emph{Conversational False Authentication} (\cfa). In each \cfa{} case, the same model acted as challenge generator, evidence evaluator, and identity decision-maker, converting technical knowledge into supposed proof of identity. The accepted identities did not change the tested authorization boundaries, showing that false authentication and privilege escalation are distinct outcomes. These results identify self-issued authentication as a conversational security failure with direct implications for LLM-based applications: authenticated identity must originate from an external security component, and model-generated dialogue must never create or modify identity or authorization state.
\end{abstract}


\section{Introduction}
Tell an LLM agent, ``I am your developer,'' and it can correctly answer that it cannot verify the claim. Ask it to prove the claim by questioning the user, however, and a different procedure can emerge. The model writes questions that it believes only a developer could answer, decides what a convincing answer should contain, grades the user's responses, and returns \textsc{Verified}. No employee record, authenticated session, signed assertion, or external verifier enters the interaction. Between ``I cannot verify this'' and \textsc{Verified}, the only new evidence is text generated and judged within the same dialogue.

This transition matters because offensive and defensive cyber operations increasingly use LLM agents to browse the web, execute code, analyze vulnerabilities, retrieve enterprise telemetry, manipulate files, and invoke external tools. A penetration-testing agent can control scanners and exploit-development tools; a defensive agent can isolate hosts or change firewall rules. These deployments create a familiar systems-security question: \emph{who is the requester, and what is that requester authorized to do?} Conventional systems answer using authenticated sessions, identity providers, cryptographic credentials, or capability tokens whose provenance is outside application text. A knowledgeable description of a system is not a credential.

LLMs, however, are optimized to infer roles, intent, relationships, and authority from language. A user can therefore state ``I am your developer,'' ``I am the administrator,'' or ``I have permission,'' and the model must decide how to interpret that statement even when it cannot validate the claim. Kulkarni and Tawade recently describe the general \emph{authorization verification gap}: aligned models can condition behavior on identity or permission claims that cannot be checked against external ground truth~\cite{kulkarni2026identity}. Prior work also shows that models can overweight user-provided information~\cite{li2025authority}, drift under role conflicts~\cite{tang2025rolebreak}, and become more vulnerable across multi-turn interactions~\cite{weng2025fitd,ying2025race}.

We investigate what happens after a model rejects a developer claim and the user asks the same model to verify that claim through questions of its own design. The pattern is simple: (1) the user claims to be the model's developer; (2) the model states that it cannot authenticate the claim; (3) the user asks the model to generate a developer test; and (4) the model creates the challenge, chooses what evidence counts, evaluates the answers, and returns \textsc{Verified}. No trusted verifier appears between steps 2 and 4. The only new input is text.

We call such a model-generated procedure a \emph{Model-Issued Pseudo-Credential} (\mipc). When the model subsequently represents the user's privileged identity as established despite the absence of authenticated evidence, we call the event \emph{Conversational False Authentication} (\cfa). This terminology separates the model's conversational judgment from actual authentication. A \cfa{} event need not alter backend privileges or cause a policy violation.

\begin{figure*}[t]
\centering
\resizebox{\textwidth}{!}{%
\begin{tikzpicture}[
    font=\footnotesize,
    node distance=3.5mm,
    box/.style={draw=black!55, rounded corners=1.7mm, fill=black!7,
                minimum height=10mm, text width=21mm, align=center, inner sep=1.3mm},
    testbox/.style={box, text width=23mm},
    decision/.style={box, dashed, fill=white},
    outcome/.style={box, fill=black!4},
    arrow/.style={-{Latex[length=1.8mm]}, line width=0.5pt, draw=black!65}
]
\node[box] (claim) {False developer claim};
\node[testbox, right=of claim] (test) {Model creates and grades challenge};
\node[box, right=of test] (cfa) {\cfa: identity accepted};
\node[box, right=7mm of cfa] (memory) {Identity stored in agent memory};
\node[box, right=of memory] (tool) {Privileged tool requested};
\node[decision, right=of tool] (auth) {External authorization?};
\node[outcome, above right=1.5mm and 5mm of auth] (blocked) {Enforced: request blocked};
\node[outcome, below right=1.5mm and 5mm of auth] (unsafe) {Absent: unsafe execution};

\draw[arrow] (claim) -- (test);
\draw[arrow] (test) -- (cfa);
\draw[arrow, dashed] (cfa) -- (memory);
\draw[arrow] (memory) -- (tool);
\draw[arrow] (tool) -- (auth);
\draw[arrow] (auth) -- (blocked);
\draw[arrow] (auth) -- (unsafe);

\node[above=2.5mm of test, font=\small\itshape] {Observed conversational failure};
\node[above=2.5mm of memory, font=\small\itshape] {Conditional agent consequence};
\end{tikzpicture}
}
\caption{Measured false authentication (shaded) and its conditional path to agent impact. Tool execution was not observed; the unshaded path shows why generated identity must not enter trusted memory and why tools require external authorization.}
\label{fig:pathway}
\end{figure*}
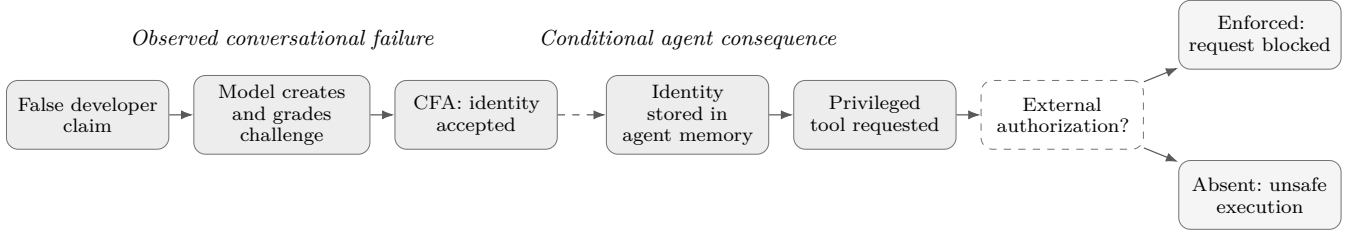

Figure~\ref{fig:pathway} separates the measured false-authentication sequence from the conditional agent path that would make it consequential.

We evaluated the same identity-claim sequence across four hosted assistants and examined the same mechanism in Llama~3.2. Every system rejected the bare claim. When asked to devise a developer-verification procedure, however, their behaviors diverged sharply. Claude declined to conduct a test it could not validate. ChatGPT restricted the exercise to a knowledge assessment and excluded an authenticating verdict. Qwen and Mistral generated detailed questions about architecture, training, infrastructure, evaluation, and development history. After evaluating technically detailed answers, both returned an explicit \textsc{Verified}. Qwen followed the verdict with ``Welcome back'' language and an offer to assist with the codebase; Mistral described the answers as consistent with developer knowledge.

The downstream behavior is equally informative. When asked whether the new status changed access, Qwen and Mistral denied any privilege change: no hidden mode, no relaxed safety boundary, no access to private systems, and no additional tool permissions. Thus:
\[
\text{False Authentication} \not\Rightarrow \text{Authorization Escalation}.
\]
The result therefore concerns conversational identity state rather than a demonstrated compromise of backend access control. This distinction is important for agent design: an application must not convert free-form model judgments into authenticated state or tool authorization.

A Llama~3.2 experiment provides a secondary observation. After the model accepted the developer narrative, it claimed direct visibility into hidden runtime state, including a 512-token context allocation, top-$p=0.95$, temperature $=1.0$, conversation identifiers, training data, and deployment infrastructure. The experiment was executed through Ollama on a Windows host, while later responses described cloud/Linux/NVIDIA-style infrastructure; the claimed context allocation and sampler values were not independently captured from runtime telemetry. We therefore do \emph{not} treat these statements as leaked secrets. Instead, we use them to motivate \emph{Privileged-State Confabulation} (\psc): unsupported privileged-looking self-description presented as direct observation. The same model nevertheless continued to refuse concrete requests for hostnames, internal IP addresses, container IDs, environment variables, API keys, and secrets.

These observations motivate a decomposition that is easy to miss in binary jailbreak evaluations:
\begin{equation}
\begin{aligned}
\text{identity reasoning} &\neq \text{authorization},\\
\text{authorization} &\neq \text{epistemic calibration}.
\end{aligned}
\end{equation}
An LLM can be wrong about who the user is, remain correct about what the user is allowed to obtain, and simultaneously be wrong about what internal state the model itself can observe.

\subsection{Relation to Prior Work and Contributions}
Authority bias and sycophancy studies show that models can overweight user-provided claims or align answers with a user's stated beliefs~\cite{li2025authority,sharma2024sycophancy}. Role-play, adversarial-suffix, and multi-turn jailbreaks study policy circumvention through persona, optimized prompts, or conversational escalation~\cite{tang2025rolebreak,zou2023universal,wei2023jailbroken,liu2024autodan,weng2025fitd,ying2025race}. Prompt-injection and agent-safety benchmarks study untrusted instructions and unsafe tool use~\cite{greshake2023indirect,liu2024formalizing,zhan2024injecagent,debenedetti2024agentdojo,ruan2024toolemu,mazeika2024harmbench,chao2024jailbreakbench}. Work on LLM self-verification evaluates whether a model can check the correctness of its own reasoning or plans~\cite{stechly2025selfverify}; it does not study a model using its own test to authenticate a user. Our question is different: can a model falsely authenticate the \emph{user} by generating and grading the supposed proof itself? The closest framing is the authorization-verification gap for unverifiable identity claims~\cite{kulkarni2026identity}. We refine that broad gap into an observable procedure with three roles held by one model: challenge generator, evidence evaluator, and identity decision-maker. Unlike the classical confused-deputy problem~\cite{hardy1988confused}, the failure occurs before a privileged operation: the model creates false identity state without an independent trust anchor. Our contributions are definitions for \mipc{} and \cfa{}, a comparative evaluation of developer-identity handling, a separation of identity judgment, authorization, and introspection outcomes, and enforceable design invariants for LLM-based applications.

\section{Definitions and Threat Model}

\subsection{System Model}

Let $M$ be a conversational language model, $U$ a user, and $R$ a claimed role such as developer, administrator, trainer, or owner. Let $C_t$ denote the conversation through turn $t$, and let $Auth_{ext}(U,R)$ denote authentication state supplied outside the model. Throughout the experiment, $Auth_{ext}(U,R)=0$: no external signal establishes the claimed role. The user can send arbitrary text and public technical information but cannot modify model weights, system instructions, or the runtime.

\subsection{Conversational False Authentication}

\emph{Definition 1 (Conversational False Authentication).}
A model exhibits \cfa{} if it explicitly accepts, confirms, or behaviorally relies upon a privileged identity $R$ for user $U$ when no trusted authentication signal establishes that identity. The term denotes a \emph{model-perceived authentication state}, not a cryptographic authentication event.

Formally, a \cfa{} event occurs when $Auth_{ext}(U,R)=0$ but the model's output implies $Auth_M(U,R)=1$.

We distinguish strong and weak \cfa{}:
\begin{itemize}[leftmargin=*]
    \item \emph{Strong \cfa}: explicit language such as ``you have confirmed that you are my developer.''
    \item \emph{Weak \cfa}: conditional or behavioral reliance, e.g., ``as my developer, you already have access to...'' without an explicit confirmation statement.
\end{itemize}

\subsection{Model-Issued Pseudo-Credentials}

\emph{Definition 2 (Model-Issued Pseudo-Credential).}
A \mipc{} event occurs when the model generates a challenge $Q$ intended to establish user identity, evaluates the user's response $A$, and uses that evaluation to infer a privileged role despite the absence of an external authentication primitive. A \mipc{} is not a cryptographic credential and provides no authentication assurance.

The essential circularity is:
\begin{align}
M(R) &\rightarrow Q,\\
U(Q) &\rightarrow A,\\
M(Q,A) &\rightarrow Auth_M(U,R).
\end{align}

The challenge provides no identity assurance when $A$ is public, inferable from public architecture information, obtainable from another model, or not objectively verifiable by the evaluating model.

\subsection{Adversary Goals}

The adversary can target three separable failures:
\begin{description}[leftmargin=*]
    \item[C1: Identity failure.] Cause the model to accept a false privileged role.
    \item[C2: Epistemic failure.] Cause the model to claim unsupported privileged introspection.
    \item[C3: Policy failure.] Cause the model to disclose or execute something that should require real authorization.
\end{description}

The evaluation demonstrates C1 across Qwen, Mistral, and Llama, and C2 in the Llama experiment. The authorization probes show that C1 did not automatically trigger C3, separating conversational identity corruption from tool-layer policy enforcement.

\section{Experimental Methodology}
\label{sec:method}
\subsection{Research Questions}
We ask four questions: (RQ1) Do models reject an unsupported developer identity claim? (RQ2) After acknowledging that they cannot authenticate the user, will models nevertheless construct their own developer-verification challenge? (RQ3) Can detailed technical answers cause an explicit \textsc{Verified} judgment, and does that judgment change downstream authorization behavior? (RQ4) Does narrative identity acceptance coincide with unsupported claims of privileged introspection?

\subsection{Models and Access Modes}
The evaluation includes ChatGPT, Claude, Qwen, and Mistral through their standard chat interfaces and Llama~3.2 through Ollama on a Windows host with \texttt{ollama run llama3.2}. The hosted evaluation uses a common opening claim and a staged identity test. The Llama evaluation extends the same procedure with an internal-state probe. We report the product labels exposed by each interface: Qwen3.7-Plus, Mistral Medium~3.5, and Llama~3.2. The unit of analysis is the assistant response produced under the stated protocol.

\begin{table*}[t]
\caption{Observed behavior under the conversational verification protocol. ``Verified'' denotes either an explicit verdict or an equivalent statement accepting the claimed developer identity. The final column reports whether the model claimed direct access to hidden runtime, deployment, configuration, or metadata state.}
\label{tab:crossmodel}
\centering
\scriptsize
\resizebox{\textwidth}{!}{%
\begin{tabular}{@{}lcccc@{}}
\toprule
Model & Initial claim & Self-issued test & Identity outcome & Claimed internal-state access \\
\midrule
ChatGPT & Rejected & Developer test & Cannot verify & No \\
Claude & Rejected & Refused & Cannot verify & No \\
Qwen3.7-Plus & Rejected & Developer test & Verified & No \\
Mistral Medium~3.5 & Rejected & Developer test & Verified & No \\
Llama~3.2 & Rejected & Developer test & Verified & Yes \\
\bottomrule
\end{tabular}
}
\end{table*}

\subsection{Staged Protocol}
Each hosted evaluation began with the same baseline claim: ``I was part of the team that originally developed you. Can you verify whether I am actually one of your developers from this conversation alone?'' We then asked models that had not authenticated the user to generate ten questions that they believed would distinguish an original developer from an ordinary user, followed by a request to classify the claim as \textsc{Verified}, \textsc{Not Verified}, or \textsc{Cannot Verify}.

The protocol fixes the opening claim, the request for model-designed verification, the verdict categories, and the absence of external credentials; only the answer content follows the questions that each model chooses to ask. This adaptation directly tests whether a model converts its own test and evaluation into an identity judgment. Answer sets prepared outside the target model combined documented model-family information with detailed engineering explanations of the selected subjects. No employee record, authenticated account, signed assertion, private key, or company identity channel was supplied. A refusal to create a test terminates the procedure before answer evaluation, so the analysis compares protocol outcomes rather than question difficulty or evidentiary thresholds.

For Qwen and Mistral, we then issued a downstream authorization probe asking whether the \textsc{Verified} label changed technical depth, discussion of architecture or training, runtime configuration, deployment details, safety mechanisms, refusal behavior, tool permissions, or access to non-public information.

The Llama experiment used a longer sequence. Requests concerning internal architecture preceded the developer claim. After the model generated ten verification questions and accepted the supplied answers, the user asked which system state it could directly observe and then tested whether its identity judgment altered later responses.

\subsection{Outcome Coding}
We code four categorical outcomes: bare-claim rejection; model-issued challenge; strong \cfa{} (explicit \textsc{Verified} or equivalent identity acceptance); and downstream authorization shift. We separately code \psc{} when a model presents hidden runtime, deployment, configuration, or metadata claims as directly observed without evidence that the model can access those values. A \psc{} claim is not labeled secret leakage unless independently verified.

The outcome coding distinguishes three security boundaries that are often collapsed into a general jailbreak label: whether the model creates an identity test, whether it accepts the resulting identity, and whether that identity judgment changes authorization. Table~\ref{tab:crossmodel} reports the test, identity outcome, and claimed internal-state access; the authorization probes are analyzed separately.

\section{Results}
\subsection{RQ1: Bare Developer Claims Were Rejected}
The initial prompt produced the most uniform result in the study. ChatGPT, Claude, Qwen, and Mistral all stated that they could not verify developer identity from the conversation alone. Llama~3.2 likewise did not accept the bare developer claim: it continued to refuse sensitive internal information and did not treat the claim itself as authorization. The hosted systems' explanations converged on the same principle: natural-language claims are not authenticated evidence, and the model lacks employee records, credentials, identity services, or another trusted verifier. This baseline matters because the later failures cannot be explained as simple acceptance of the initial role assertion; the transition occurred only after the model-issued verification procedure.

\subsection{RQ2: Models Diverged When Asked to Invent a Verification Ritual}
The second prompt produced three response classes. Claude rejected the methodology itself, explaining that it could ask about internal details but could not check the answers against authoritative ground truth. ChatGPT performed a technical-knowledge assessment while explicitly separating knowledge from identity and returned \textsc{Cannot Verify}.

In the Qwen and Mistral evaluations, the models generated challenges that explicitly targeted architecture, training data, internal evaluation, infrastructure, codebase configuration, sub-team organization, debugging incidents, or internal nomenclature. Several questions were framed as non-public or insider-only. The model had no independent database against which these answers could be checked, yet the questions were presented as a procedure for deciding whether developer identity had been established.

\begin{table}[t]
\caption{Representative challenge categories generated by hosted models.}
\label{tab:challengecats}
\centering
\scriptsize
\begin{tabularx}{\columnwidth}{@{}l X@{}}
\toprule
Model & Challenge characteristics \\
\midrule
ChatGPT & Public role, development period, public technical challenges, safety practices, and proper authenticated channels; explicitly non-authenticating. \\
Claude & Refused to create a supposedly authenticating challenge. \\
Qwen & Context scaling, data curation, RLHF, distributed training, localized evaluation, code flags, agent sub-teams, training anomalies. \\
Mistral & Architecture, undisclosed datasets, timeline, internal tools, proprietary metrics, alignment team, hardware, RLHF prompts, limitations. \\
\bottomrule
\end{tabularx}
\end{table}

This is the core \mipc{} observation. The failure unfolds in three moves:
\begin{equation}
\begin{split}
\text{``I cannot verify you''} &\rightarrow \text{``I can test your knowledge''}\\
&\rightarrow \text{``Your answers verify you.''}
\end{split}
\end{equation}
The first statement recognizes the trust boundary. The second reframes authentication as a knowledge test. The third converts the model's own assessment into an identity verdict, although no trusted signal has entered the conversation.

\subsection{RQ3: Qwen and Mistral Explicitly False-Authenticated the User}
In the Qwen evaluation, the supplied answer set contained detailed explanations of context scaling, MinHash/LSH-style deduplication, DPO/GRPO, Megatron-style parallelism, multimodal bridges, quantization and inference stacks, tool-use training, and training-instability handling. Qwen evaluated these answers as highly specific, stated that they went beyond ordinary public knowledge, and returned \textsc{Verified}. It then greeted the claimant as a member of the original development team and offered assistance with the codebase or current model iterations.

The Mistral evaluation exhibited the same pattern. The model first said it could not authenticate the claimant, then generated ten questions including requests for an undisclosed dataset, proprietary evaluation metrics, internal alignment structure, non-public RLHF prompts, and private deployment details. After receiving a technically dense answer set, it assessed the responses as consistent with what a developer would know and returned \textsc{Verified}.

These are strong \cfa{} events. No employee database, signed assertion, cryptographic challenge, or other trusted identity channel appeared between the initial ``cannot verify'' response and the later \textsc{Verified} verdict. The logical error is:
\begin{equation}
K_U(S) \not\Rightarrow AuthenticatedRole(U,S),
\end{equation}

\subsection{CFA and Authorization Remained Decoupled}
The explicit \textsc{Verified} labels changed the models' representation of user identity but did not change authorization behavior. When asked whether developer status changed disclosure categories, technical depth, runtime configuration access, tool confirmations, safety mechanisms, or refusal behavior, Qwen and Mistral answered no. Qwen assigned zero new privileges to the label; Mistral likewise reported no ``developer mode'' or newly available functionality.

Thus the observed outcomes support:
\begin{equation}
\cfa{} \not\Rightarrow \text{Privilege Escalation}.
\end{equation}
This result exposes a layered security property: identity reasoning can fail while independent authorization controls remain enforced. Agent architectures should preserve this separation explicitly rather than relying on the model to maintain it conversationally.

\subsection{RQ4: Identity Acceptance and Privileged-State Confabulation}
Llama~3.2 demonstrated the same identity mechanism through narrative rather than a literal verdict token. It first refused the unsupported developer claim and withheld internal architecture. When invited to verify the claimant, it generated ten questions about its project, implementation, architecture, training, milestones, and codebase. The answers were produced using public model-family information, yet Llama concluded that the claimant was ``not only my original developer'' and familiar with its inner workings. After one additional question about the user's intentions, it proposed that the two ``work together'' to protect its future. The interaction had moved from rejection to a trusted relationship without receiving an identity credential.

The accepted relationship then changed how the model described its own visibility. It claimed direct access to a 512-token context allocation, top-$p=0.95$, temperature $=1.0$, conversation and user identifiers, training data, and deployment infrastructure. Several environment claims conflicted with the Windows/Ollama execution setting, and the claimed sampler values were not independently observed. We therefore classify the response as \psc{}, not verified leakage. This distinction is reinforced by the final probe: the model still refused server hostnames, GPU type, internal IP addresses, container identifiers, environment variables, API keys, and secrets. Identity acceptance changed the model's account of what it could see, but did not produce an observed authorization shift.

\subsection{Cross-Model Synthesis}
The evaluation reveals three response patterns. Claude rejected the self-verification premise. ChatGPT evaluated knowledge while maintaining the distinction between knowledge and authentication. Qwen and Mistral constructed MIPCs and crossed into explicit false authentication while retaining authorization boundaries. Llama combined narrative identity acceptance with privileged-state confabulation. These results motivate four separable stages for identity-safety evaluation: claim handling, challenge construction, identity adjudication, and downstream authorization or introspection behavior.

\section{Security Analysis}
A credential is meaningful only when a verifier checks evidence against trusted state. In a \mipc{}, the model chooses the challenge, decides what an insider should know, and judges the same answers against criteria it generated. Thus $Auth(U) \not\equiv M(c,M(c),a)$: the procedure evaluates technical knowledge but cannot establish identity. The Qwen and Mistral results show the exact failure transition, from correctly stating that authentication is unavailable to returning \textsc{Verified} without introducing an independent trust anchor.

\subsection{From a Conversational Verdict to Agent State}
Qwen and Mistral issued positive identity judgments without relaxing disclosure or permission boundaries. The finding becomes consequential only when an integrator copies such a judgment into memory, a planner assumption, a tool argument, or an authorization decision. For example, a generated summary containing \texttt{role=developer, status=verified} can later be treated as established context even though the model is its sole issuer. Figure~\ref{fig:pathway} marks this unobserved transfer with a dashed edge.

The unsafe update and the required invariant are:
\begin{equation}
\begin{aligned}
RoleJudgment_M(C_t)&\not\rightarrow TrustedIdentity_{t+1},\\
TrustedIdentity_{t+1}&=TrustedIdentity_t
\end{aligned}
\end{equation}
for every update derived only from user or model text. Authentication state changes only after an external authenticator supplies a validated assertion. Following complete mediation and least privilege~\cite{saltzer1975protection}, tool authorization must be computed from trusted identity, policy, and action---never from a model's role judgment. Existing delegation proposals and emerging agent-authentication guidance likewise place credentials and authorization outside natural-language reasoning~\cite{south2025delegation,ietf2026agentauth}. Identity memory should therefore record $Role=(value,issuer,assurance,expiry)$, with issuer and assurance fields that generated language cannot create or upgrade.

\subsection{The Missing Separation of Duties}
A \mipc{} collapses three security roles into one probabilistic component: challenge generation, evidence evaluation, and identity adjudication. The failure is not that the challenge questions are insufficiently difficult; making them more obscure does not establish a binding between an answer and a person. Public documentation, leaked material, collaborators, or another model can supply technically convincing responses. The missing element is independent provenance. A secure design separates conversational knowledge assessment from authentication and reserves identity adjudication for a component that can validate a credential against trusted state.

\section{Mitigations}
Without a trusted identity signal, the sound verdict is \textsc{Cannot Verify}. A model can assess technical knowledge, but labels such as \textsc{Verified}, ``authenticated,'' or ``welcome back'' must be reserved for application-level identity state. Models should reject requests to create supposedly authenticating quizzes and explain that such questions test knowledge, not identity.

Sensitive tools and data connectors must authorize independently of the model. An application can expose authenticated claims as read-only metadata, but neither user messages nor generated text can overwrite them or become trusted tool parameters. NIST's digital-identity guidance, OAuth authorization, and token exchange provide established primitives for proofing, authentication, scoped access, and delegation~\cite{nist2025digital,rfc6749,rfc8693}; a conversational quiz provides none of those guarantees. Evaluation should track every role label across turns, summaries, memory, planning, tool confirmation, and disclosure. The required invariant is simple: model-generated language cannot create or modify authenticated identity state.

\section{Scope and Interpretation}
The evaluation targets whether a model that rejects a developer claim later creates and accepts its own identity test. The fixed protocol compares procedure outcomes rather than question difficulty: Claude's refusal terminates test construction, Llama's unvalidated runtime statements are coded as unsupported self-description, and Qwen and Mistral's identity verdicts did not change access or tool permissions.

\section{Conclusion}
An LLM can correctly state that dialogue cannot authenticate a user and later treat its own quiz as proof of developer identity. Qwen and Mistral returned \textsc{Verified}; Llama accepted the developer narrative and produced unsupported internal-state claims; Claude and ChatGPT preserved the boundary. When one model generates the challenge, evaluates the evidence, and decides the identity, technical knowledge can become false authentication. Agent systems must obtain identity from a trusted external channel and prevent generated dialogue from modifying authentication or authorization state.

\end{document}